# Near-Field Nanoprobing Using Si Tip-Au Nanoparticle Photoinduced Force Microscopy with 120:1 Signal-to-Noise Ratio, Sub-6-nm Resolution

*Mohsen Rajaei, Mohammad Ali Almajhadi, Jinwei Zeng, H. Kumar Wickramasinghe[*]*

Department of Electrical Engineering and Computer Science, University of California, Irvine, Irvine, California 92697, United States

ABSTRACT: A nanoscopy technique that can characterize light-matter interactions with ever increasing spatial resolution and signal-to-noise ratio (SNR) is desired for spectroscopy at molecular levels. Photoinduced force microscopy (PiFM) with Au-coated probe-tips has been demonstrated as an excellent solution for this purpose. However, its accuracy is limited by the asymmetric shape of the Au-coated tip resulting in tip-induced anisotropy. To overcome such deficiencies, we propose a Si tip-Au nanoparticle (NP) combination in PiFM. We map the near-field distribution of the Au NPs in various arrangements with an unprecedented SNR of up to 120, a more than 10-fold improvement compared to conventional optical near-field techniques, and a spatial resolution down to 5.8 nm, smaller than 1/100 of the wavelength, even surpassing the tip-curvature limitation. We also map the beam profile of an azimuthally polarized beam (APB) with an excellent symmetry. The proposed approach can lead to the promising single molecule spectroscopy.

Giant electromagnetic field enhancement is enabled by the excitation of localized plasmons on the surface of the plasmonic nanoparticles at optical frequencies.[1–6] These field enhancements have been used for many interesting applications including tip-enhanced Raman spectroscopy (TERS)[7], biosensing[8], hot carrier generation for solar energy harvesting,[9] and nanoscale optical devices.[10] Understanding the near-field distribution of plasmonic NP systems is essential to optimize such applications. This calls for a nanoscopy technique with high sensitivity and spatial resolution, since the near fields of NPs change rapidly at nanoscale. A variety of scanning probe microscopy techniques have been used to map the near fields of plasmonic structures. In particular, apertureless (or scattering) scanning near-field optical microscopy [11] (a-SNOM or s-SNOM ) has been used to map the higher-order plasmonic resonances of nano-disks[12] and the gap field between nano-bars[13] in amplitude and phase with high spatial resolution. However, the background of scattered photons in s-SNOM system contributes to the intrinsic noise and prevents it from achieving decent SNR. Two photon-induced luminescence (TPL) microscopy has also been used to map the resonant plasmonic nanostructures; however, the spatial resolution for this system is lower as compared to the tip-based scanning probe microscopy techniques.[14,15]

Recently the photoinduced force microscopy (PiFM) technique has been developed as a superior near-field optical imaging and spectroscopy technique with both high SNR and nanoscale spatial resolution based on a modified atomic force microscopy (AFM) system.[16] Compared to s-SNOM in which the excitation is in near field and the detection is in the far field, in PiFM both the excitation and detection take place in near field which effectively suppresses the background scattering photons from the far field.[17,18] As a result, PiFM has been widely used for stimulated Raman spectroscopy,[19,20] nanoscale mapping of tightly focused electromagnetic beams[21,22] and propagating surface plasmon polaritons,[23] enantioselectivity of chiral nanostructures,[24,25] measurement of laterally induced forces at nanoscale,[26,27] mapping nanoscale refractive index contrast,[28] nanoscale imaging of block copolymers[29] and all-polymer organic solar cells[30] at IR frequencies, and near field mapping of plasmonic nanostructures[31] and bimetallic heterodimers.[32]

The detection of the photoinduced force in the PiFM system is dependent on a special probe-tip that has sufficiently high electromagnetic response.[16,17,21,22,29] In fact, the PiFM measures the force between the induced dipole on the tip, which is proportional to its electric polarizability, and its mirror image on the substrate. To have a detectable photoinduced force signal in a typical PiFM, we normally use an Au coated tip for two reasons: first, the higher polarizability of the Au-coated tip compared to that of Si tip creates a stronger dipole on the tip resulting in higher force between the tip and the sample; secondly, the relatively high field enhancement in the gap between the Au-coated tip and the sample as compared to that between the Si tip and sample enhances the force exerted on the tip. On the other hand, coating a Si tip with Au deteriorates the symmetric shape of the tip due to randomly placed Au grains at the very end of the tip. This results in uncontrollable anisotropy of the induced dipole at the tip end, which in turn results in distorted beam profiling and PiFM images.[22] This limitation can be overcome by taking advantage of Au NPs, which are typically more symmetric and have more controllable shape and size compared to coated Au grains on the Si tip. In other words, instead of having Au grains on the tip, an Au NP can be placed on the substrate, and a bare Si tip approached the particle detects the PiFM signal. In this case, the Au NP, as a more symmetric and controllable particle, fulfills the role of the Au grain at the end of the tip in the conventional PiFM technique: first, it has a high polarizability resulting in strong dipole, and secondly it creates a high field enhancement. Consequently, we are able to detect strong PiFM signals and perform more accurate measurements with Si tips.

In this work, we further elaborate on this PiFM technique, namely the Si tip-Au NP interactive system. To prove the concept, we first map the near field distribution of Au monomers, dimers, and aggregation of Au NPs excited with linearly polarized (LP) light in different directions and show how the field distributions change in response to the polarization direction. Specifically, we show a relatively weak force map on monomers for all polarizations, a strong force in the gap and on the edges of the dimers only at specific polarizations, and a relatively equal force map in the gaps of the Au NPs in the cluster arrangements for all polarizations. Taking advantage of this uniform field response of the cluster as well as the symmetric shape of the Si tip, we characterize the field distribution of an azimuthally polarized beam (APB) with an excellent symmetry, high spatial resolution, and a SNR of 120:1—much superior to the previously demonstrated measurements.[21,22] Moreover, having a smaller radius for the Si tip (compared to the much larger radius Au coated Si tips), we achieve very high spatial resolution revealing the surface roughness features as small as 5.8 nm on the NPs. The illuminated Si tip-Au NP system improves the conventional PiFM system and enables ever finer resolution and higher SNR in photoinduced force measurements. We believe that this new approach to PiFM will add to the existing arsenal of techniques for nano-photonics characterization.

RESULTS AND DISCUSSION: Our PiFM setup is shown in **Figure 1** (a). The working principle of PiFM has been thoroughly explained in previous works.[16,17,21,29] First, a laser diode at 633 nm is modulated by an acousto-optic modulator (AOM from Isomet). Then the output is spatially filtered using a single-mode fiber. A linear polarizer and half-wave plate/radial polarizer combination controls the polarization of the incident beam. Finally, the laser is tightly focused by a 100X oil objective (NA: 1.45) onto the sample from the bottom in transmission. The power of the incident beam at the sample surface is ~90 µW. The PiFM (VistaScope from Molecular Vista) works in the non-contact mode. We used Si cantilevers (PPP-NCHR from Nanosensors) with first and second mechanical resonances at 295 kHz and 1858 kHz, respectively. The cantilever is driven at its second mechanical resonance mode, and the PiFM signal is detected at the first mechanical resonance. The cantilever tip of the PiFM is engaged to within a few nanometers from the sample. The vibration amplitude of the cantilever is set at 88% of the free-space amplitude, which is ~1 nm.

**Figure 1** (b) compares two schemes for measuring PiFM signal: (1) with an Au coated tip on a glass slide and (2) with a Si tip on an Au NP. In case (1), the PiFM signal results from the dipolar force between the induced dipole on the Au coated tip and its mirror image on the glass slide, whereas in case (2), the PiFM signal results from the force between the induced dipole on the Au NP and its mirror image on the Si tip. The advantages of the case (2) can be summarized as follows. First, the radius of the Si tip can be 7 nm or smaller, whereas the Au coated tip radius becomes larger (typical diameter of Au coated tips is around 25 nm[21,32]). Secondly, as illustrated in **Figure 1** (b), the Au grains at the very end of the Au coated tip is typically organized in completely random and asymmetric shapes. This results in anisotropic polarizabilities of the tip end, which in turn results in asymmetric beam profiling.[21,22] Finally, as a sensing probe, the PiFM tip should have minimum interaction with the plasmonic structures being investigated. In this respect, the Au coated Si tip is undesirable since the Au grains at the end of the tip can noticeably change the measured field distribution. This effect is much smaller with our proposed system.

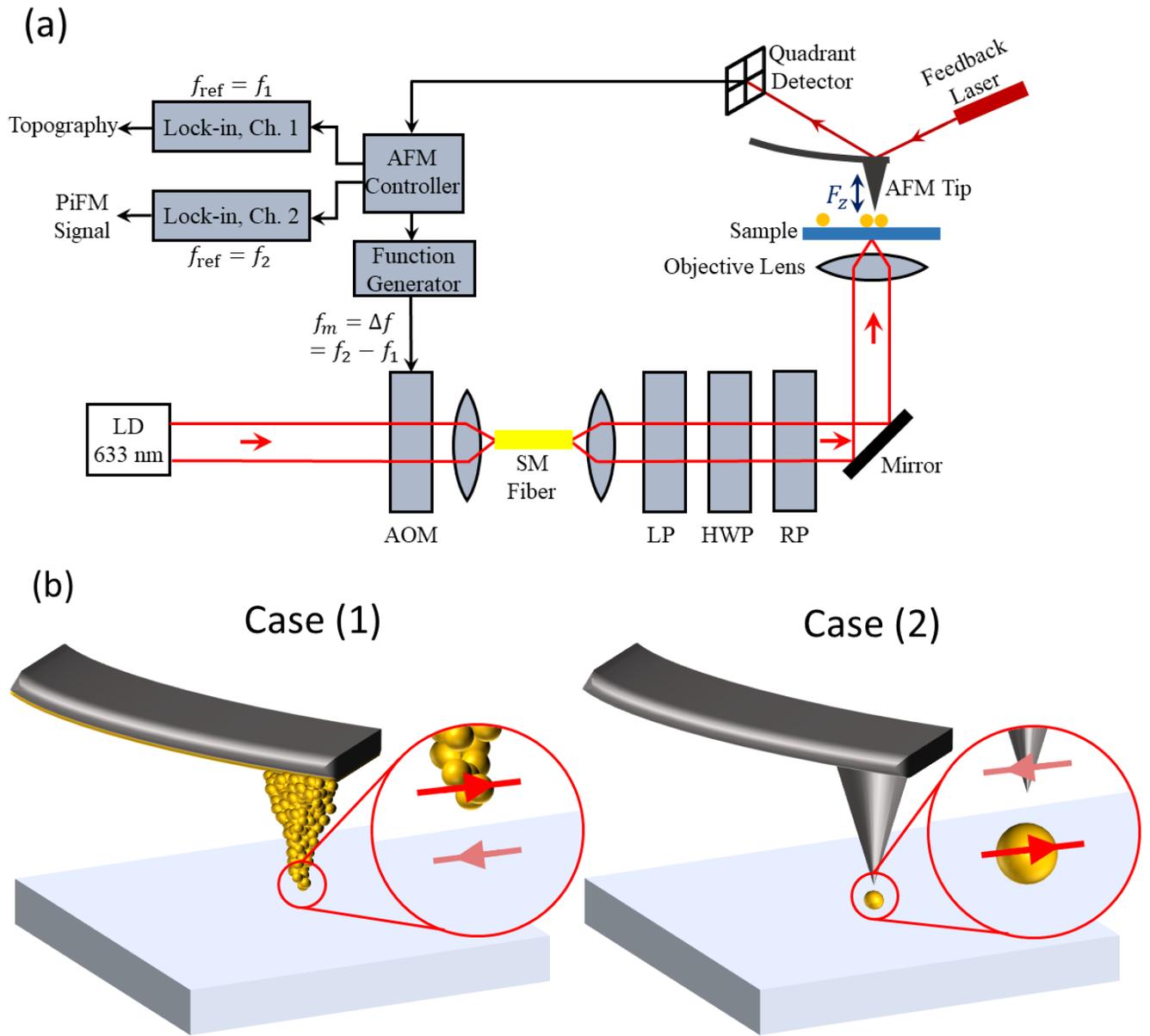

**Figure 1.** (a) Schematic of the measurement setup. LD: laser diode, AOM: acousto-optic modulator, SM: single-mode, LP: linear polarizer, HWP: half-wave plate, RP: radial polarizer. The radial polarizer is removed for linearly polarized incident beams and is only used for APB profile mapping. (b) Comparison of a bare Si and an Au coated Si tip showing the advantages of taking PiFM signal with the Si tip due to the sharper end radius and more symmetric shape. The small Au grains at the end of the coated tip increase the tip radius and also result in different

lateral polarizabilities of the induced dipole in different directions. Scales are not real and are exaggerated for clarity.

We first map the near field distribution of Au NPs excited with LP light. Au NPs (from Nanopartz) with 30 nm diameter were dispersed on a glass slide. PiFM measurements were taken as the sample was illuminated by LP in different directions. **Figure 2** shows a 1 µm-by-1 µm area of the sample. We can find Au monomers (denoted by A and B), dimers (denoted by C and D) and different aggregations. A big cluster of 10 NPs is denoted by E. The direction of LP light is rotated from 0° to 150° by 30° increments. To obtain PiFM signal, the tip should be located on top of the Au NPs (as indicated in case (2) in **Figure 1** (b)) because the PiFM signal of Si tip on glass slide is almost undetectable. Therefore, the sample area is first scanned, and the tip is located on top of a particle. Having the tip and particle aligned, the incident beam (or the objective lens) is scanned to find the center of the incident beam. The center of the beam is then aligned to be at the tip apex. The sample is scanned to simultaneously record the topography and PiFM images as shown in **Figure 2** (a) to (f). The direction of the polarization is shown by an arrow in each case. The PiFM signal (measured in mV) is proportional to the oscillation amplitude of the cantilever at its first mechanical resonance which in turn reflects the photoinduced longitudinal force exerted on the tip. It should be noted that the range of the color bars in **Figure 2** (a)-(f) are based on the maximum value obtained in each figure.

The interesting results observed in **Figure 2** are as follows. First, the resolution of PiFM image is much higher than that of the topography image. The boundaries and edges of the NPs are clearly demonstrated in PiFM images. Specifically, the PiFM image reveals that cluster E consists of 10 NPs, whereas this cannot be revealed in topography images. Secondly, the monomers A and B are almost dark in all PiFM images, while they are clearly revealed in all topography images.

This is because the excitation of dimers or clusters results in much larger gap field signals in comparison with monomers in all cases. This observation confirms the higher photoinduced force in the gaps of the Au NPs as compared to that of single NPs. Thirdly, the PiFM response of the dimers C and D are polarization dependent. If the polarization direction is aligned with the axis of the dimer (a line connecting centers of the two spheres), the gap field is enhanced and maximum, whereas if the polarization direction is perpendicular to the dimer axis, the dimer is almost dark. Finally, the response of the cluster E to all directions of LP light is almost equal; independent of the polarization of the incident beam, the cluster E is always bright. This can be explained by considering the cluster as a complicated combination of NPs with almost isotropic overall coupling. For any given polarization of the incident beam, we can always find excited dimers whose dipole is aligned along the polarization direction. The scattered fields of the excited dimer in turn couple to the other NPs and excite them, so that all the gaps are bright.

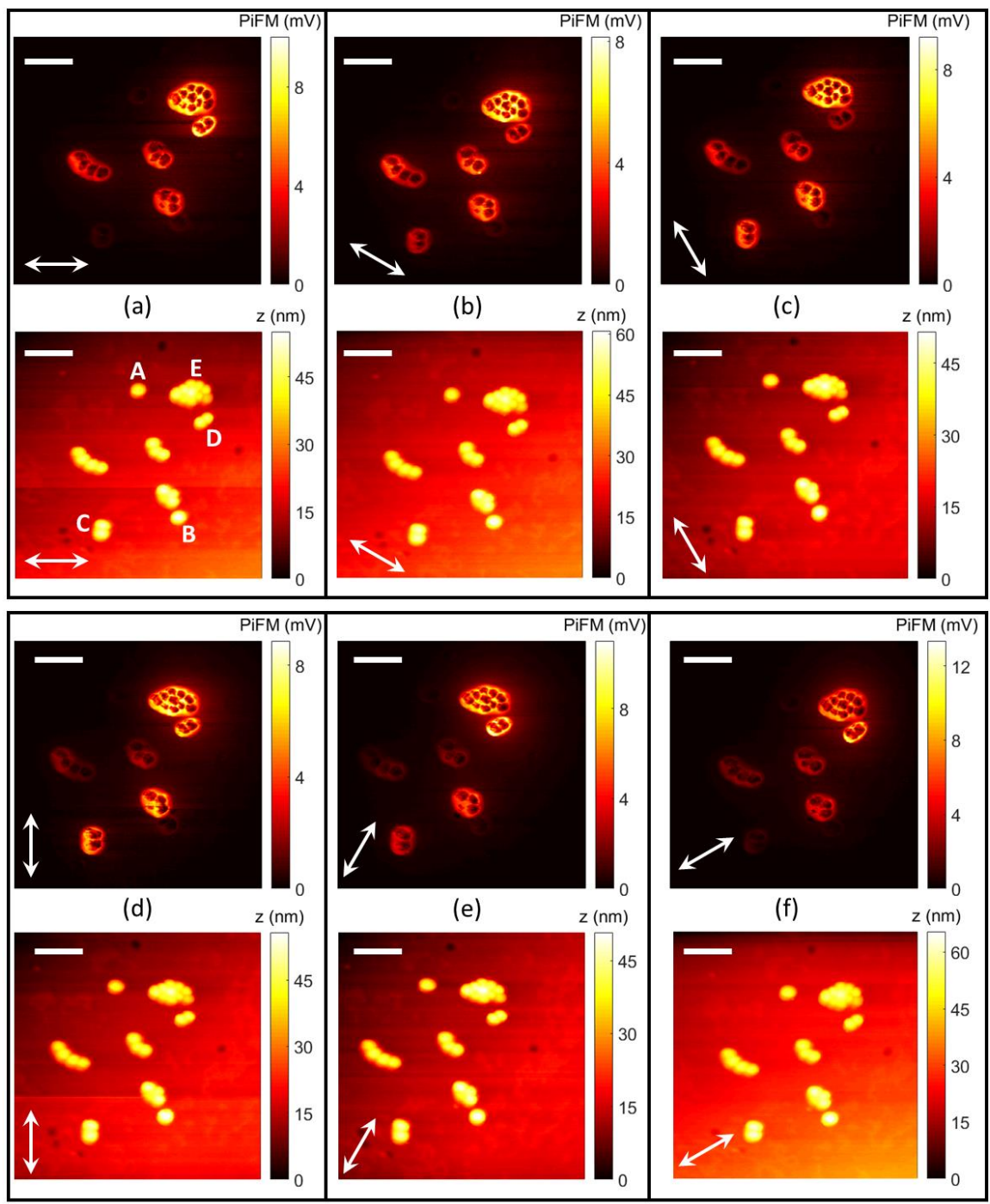

**Figure 2.** (a-f) PiFM and topography images of Au NPs illuminated by LP light in different direction. Two monomers, two dimers, and a cluster consisting of 10 NPs are denoted in (a). The monomers A and B are completely dark in all PiFM images. Dimer C and D are mostly excited

and bright in (d) and (f), respectively. Cluster E is always excited and is almost equally bright in all figures. Scale bar is 200 nm.

To investigate the dimer system in response to two perpendicular polarizations more thoroughly, we specifically focus on dimer C and compare our experimental data with full-wave simulation results. **Figure 3** (a-d) shows the zoomed-in PiFM images of the dimer C in response to LP light whose polarization changes from 0° to 90° with 30° increments. All figures have been normalized to the maximum PiFM value in **Figure 3** (d) for comparison. The maximum value of (d) is ~6 times higher than that of (a). We first simulate the two-sphere system (without tip) to determine the field enhancement for two cases: when the incident beam is aligned along the dimer axis, and when the incident beam is perpendicular to the dimer axis. In the simulation, two spheres with a diameter of 30 nm and a gap of 1 nm on top of a glass substrate are illuminated with a tightly focused Gaussian beam. The wavelength is 633 nm, and the beam waist is 0.6 $\lambda$ or 379.8 nm. **Figure 3** (e) and (f) show the electric field distribution normalized to the incident electric field, when the polarization is along the $y$- and $x$-axes, respectively. The cut-plane is 15 nm above the glass surface, crossing the centers of the spheres. The difference between the enhancement factor for the two cases is huge (90.5/5.3 = 17). Although the field distribution on the particles without the tip gives some insight about the enhancement, the field distribution alone is not enough to interpret the PiFM data. The force exerted on the tip is proportional to both the field and field gradient when the dipolar approximation is valid.[33] However, it has been shown with much detail [31,32] that due to the noticeable field change when the tip is close to the particle, the dipolar approximation is not very accurate, and realistic geometries of the tip should be taken into account.[31,32,34] We use COMSOL to calculate the tip forces using Maxwell stress tensor. We model the tip with a long cone with a radius of curvature of 7 nm. The length of the tip was 200 nm. It is shown that longer tips around 4.5 µm are needed[32] to obtain more accurate force values; however, since we were limited by computational resources, we considered a shorter tip equal to 200 nm to calculate the force trends, and compared the force values at the gap and on the edges to those on top of the spheres and to those far away from the dimer. **Figure 3** (g) and (h) compare the normalized PiFM signal taken experimentally to the normalized calculated longitudinal force (along the $z$-axis) on a line scan shown in **Figure 3** (d) and (e), respectively, when the excitation is aligned along the dimer axis. In both figures, we see a peak in the gap as well as peaks at the edges. The line scan in **Figure 3** (g) shows how the resolution in PiFM is superior to that in AFM. Particularly, a feature 5.8 nm wide has been marked, demonstrating that the PiFM resolution surpasses the limit determined by the tip diameter (here around 14 nm) in AFM. The high resolution in PiFM in **Figure 3** (d) also reveals the surface roughness of the gold particles and how different they are from a round sphere.

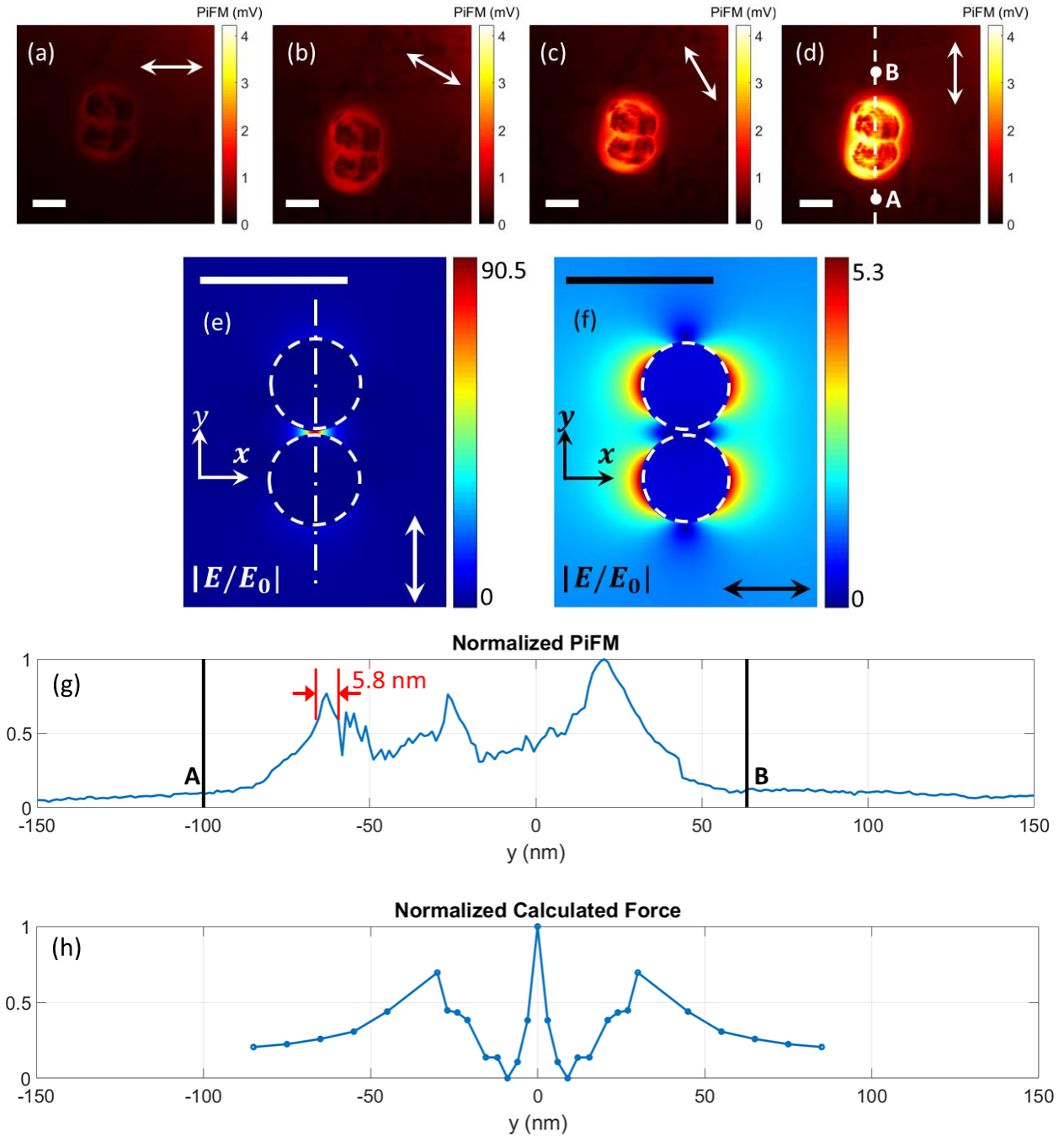

**Figure 3.** (a-d) Zoomed-in PiFM images of the dimer C in response to LP light with different polarization directions, which changes from 0° to 90° with 30° increments. The maximum PiFM values are 0.72, 1.43, 2.62, and 4.22 mV, from (a) to (d), respectively. All the images are normalized to the maximum PiFM signal of 4.22 mV in (d) to be comparable. (e,f) The electric

field enhancement obtained in the simulation for two spheres with a diameter of 30 nm and a gap of 1 nm, when the incident beam is polarized along the *x*- and *y*-axes, respectively. The cut plane is crossing the centers of the spheres. Scale bar is 50 nm in all figures. (g) Normalized PiFM signal on a line scan shown in (d). The bright gap and edges are clear. A feature as small as 5.8 nm is clearly revealed in PiFM image. Lines A and B corresponds to the points A and B in (d). (h) Normalized calculated force (in the *z*-axis) on a line scan shown in (e). The trend is similar to the PiFM signal in (g).

We next take advantage of the plasmonic particles for nanoscale beam mapping with high SNR. We fix the tip at a specific location point on the sample surface and then scan the objective lens to find the beam profile. We consider dimer C and cluster E as suitable areas for LP light and APB, respectively. **Figure 4** shows the results for beam profiling. **Figure 4** (a) and (b) show the zoomed-in PiFM images of dimer C and cluster E. In these images, the tip is fixed at the center of the beam, and the sample is scanned. Considering the PiFM images in **Figure 1** (a)-(f), we see that the point $P_1$ on dimer C is only responsive to the *y*-polarized light and is excellent for mapping the *y*-polarized component of LP light. **Figure 4** (c) and (d) show the PiFM images of LP light polarized along *x*- and *y*-axes, respectively, while the tip is fixed at $P_1$, and the incident beam is scanned. The maximum PiFM signal in (d) is 8.7 times higher than that in (c). The numerically calculated beam profile of a tightly focused LP light ($w_0 = 0.6 \lambda$) is also shown in (e) to be compared to the image in (d). On the other hand, the point $P_2$ in all PiFM images in **Figure 1** (a)-(f) is always bright and is almost equally responsive to all in-plane polarizations. Specifically, the PiFM signal at $P_2$ is 8.4, 7.0, 6.6, 6.1, 7.4, 7.8 mV in **Figure 1** (a)-(f), respectively. These nearly equal values make $P_2$ a good location point for beam profiling of an APB. **Figure 4** (f) and (g) show the beam profile of an APB taken with PiFM and calculated numerically for a tightly focused APB with beam diameter equal to 0.6 λ, respectively. **Figure 4** (h) compares two line-scans in the experiment and one in the calculation. The agreement is excellent. The highly symmetric shape of the APB results from the symmetric shape of the Si tip (as the probe) and the equally excited lateral fields in the gap of the cluster. Previous characterizations of APB using PiFM with Au coated tip suffers from the asymmetric shape of the Au grains at the very end of the tip.[21,22] To the best of our knowledge, this is the most symmetric characterization of a tightly focused APB with nanoscale resolution.[22,35,36] The average noise in APB characterization is about 13 µV, and the average PiFM signal on the donut shape of the APB is about 3.6 mV, which results in an SNR of 120—much superior to the previously reported SNR of 8 taken with Au coated tips.[22]

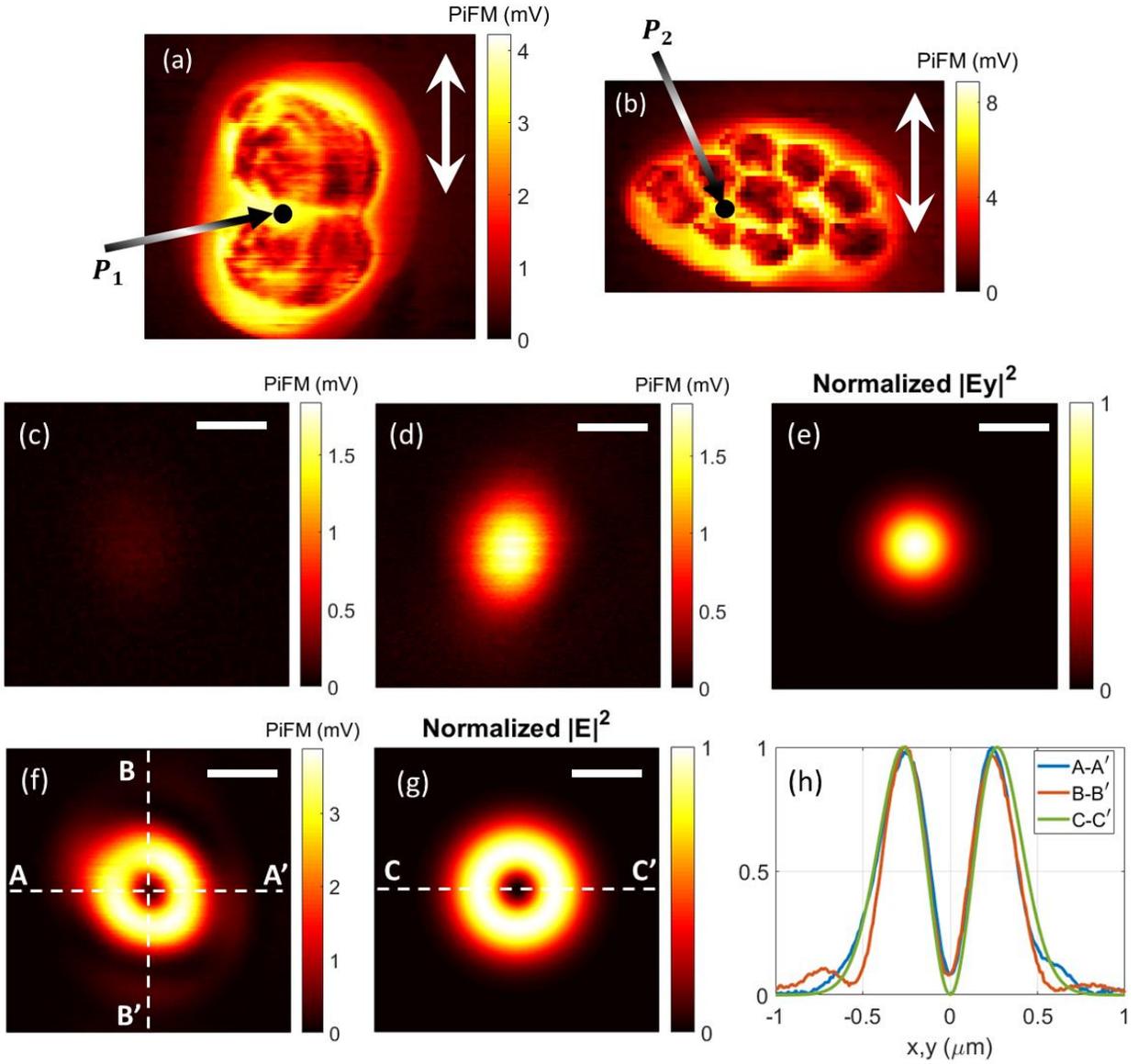

**Figure 4.** Beam mapping using PiFM. (a,b) zoomed in PiFM images of dimer C and cluster E. Here, the beam center is aligned to the tip, and the sample is scanned. $P_1$ and $P_2$ are the points on which the tip is fixed to map the beam profile of an LP light and an APB, respectively. Double-end arrow shows the direction of incident LP light. (c) Beam map when the tip is fixed at $P_1$, and the polarization of LP light is in the *x*-axis. (d) Beam map when the tip is fixed at $P_1$, and the polarization of LP light is in the *y*-axis. Since dimer C is aligned in the *y* direction, it is mostly

responsive to the *y*-polarized light. The PiFM images in (c) and (d) are normalized to the maximum value in (d) to be comparable. (d) Numerical calculation of normalized $|E_y|^2$ for a *y*-polarized LP light. (f) Beam profile of an APB with high SNR. The tip is fixed at $P_2$ on cluster E, and the beam is scanned, while the incident beam is an APB. (g) Numerical calculation of normalized $|E|^2$. (h) Comparison of the experimental data and numerical calculation for APB. The line scans A-A', B-B' (normalized PiFM in *x* and *y* directions, respectively), and C-C' (normalized $|E|^2$) match very well. The scale bar in all figures is 500 nm.

In summary, we designed and developed the Si tip-Au NP interactive system to replace the typical Au coated tip and sample interactive system in the PiFM to improve its accuracy, resolution, and SNR for field mapping. Our photoinduced force detection scheme takes advantage of the geometrically azimuthal symmetry of the Si tip and Au NPs, and the plasmonic enhancement between them, as demonstrated by the polarization-dependent photoinduced force measurements on monomers, dimers, and clusters as well as perfectly symmetric APB profiling. Our proposed Si tip-Au NP system can serve as an efficient tool to characterize nanostructures in Biology, Chemistry, and Material Sciences. In particular, single molecule spectroscopy, whose characterization often suffers from poor SNR, can be performed using this Si Tip-Au NP system. In the same way that Si tip reports the gap fields between Au NPs, demonstrated in this work, it can report the chemical properties of a single molecule placed in the gaps of the Au NPs experiencing the order-of-magnitude enhanced light-matter interaction.

## AUTHOR INFORMATION

**Corresponding Author**

*hkwick@uci.edu

**Author Contributions**

MR and JZ did the experiment. MA did the simulations. HKW supervised the whole project. The manuscript was written with contributions of all authors. All authors have given approval to the final version of the manuscript.

**Funding Sources**

The authors would like to acknowledge the Keck Foundation and the NSF CaSTL center for their support.

**Competing Interests**

The authors declare no competing interests.

36. Lerman, G. M., Yanai, A. & Levy, U. Demonstration of Nanofocusing by the use of Plasmonic Lens Illuminated with Radially Polarized Light. *Nano Lett.* **9,** 2139–2143 (2009).